\documentstyle[aps]{revtex}

\begin{document}
\twocolumn [\hsize\textwidth\columnwidth\hsize
           \csname @twocolumnfalse\endcsname
\draft
\title{Zipf's Law in the Liquid Gas Phase Transition of Nuclei}
\author{Y. G. Ma}
\address{
CCAST (World Laboratory), P. O. Box 8730, Beijing 100080, CHINA\\
 Shanghai Institute of Nuclear Research, Chinese Academy of 
Sciences, P.O. Box 800-204, Shanghai 201800, CHINA\\
}
\date{Published in Euro. Phys. J. A {\bf 6} (1999) 367-371 as a short note}
\maketitle
\begin{abstract}
Zipf's law in the field of linguistics is tested in the nuclear 
disassembly within the framework of  isospin dependent lattice 
gas model. It is found that the average cluster charge (or mass)  
of  rank $n$ in the charge (or mass) list shows exactly inversely 
to its rank, i.e., there exists Zipf's law, at the phase transition  
temperature. This novel criterion shall be helpful to search 
the nuclear liquid gas phase transition experimentally and 
theoretically. In addition, the finite size scaling of the 
effective phase transition temperature at which the Zipf's law 
appears is studied for several systems with different mass 
and the critical exponents of $\nu$ and $\beta$ are tentatively
extracted. 
\end{abstract}
\vskip2pc]
\vspace{1.cm}

\narrowtext

In intermediate  energy heavy ion collisions (HIC)  the hot 
nuclei with moderate temperature can be formed and  finally 
de-excite by different decay modes, such as multifragmentation. 
This kind of multifragment emission shows a rise and fall with  
beam energy (excitation energy, or nuclear temperature, or 
impact parameter ...) \cite{Maprc95}, which  probably relates 
to the critical point behavior in nuclear matter. The onset of 
multifragmentation, i.e., when the multiplicity $N_{imf}$ of 
intermediate mass fragment (IMF) attains to the maximum, probably 
indicates of the coexistence of liquid and gas phases. Relations 
to the IMF distribution, theory predicted that there exists a 
minimum of power law parameter, $\tau _{min}$, for the IMF cluster 
distribution when the liquid gas phase transition takes place 
\cite{Fisher}. 

On the other hand, the study on nuclear caloric curve is an 
attractive subject due to the fact that a temperature plateau 
exists when the first order liquid gas phase transition occurs.  
Experimentally He-Li isotopic temperature from Albergo's 
thermometer \cite{Albergo} for projectile-like Au spectators 
seems to have a temperature plateau in the excitation energy 
range of 3 to 10 MeV/u by ALADIN Collaborations \cite{Aladin}, 
which was claimed that the first order nuclear liquid gas phase 
transition was observed experimentally. 
 Nuclear caloric curves were also surveyed by several groups, 
such as INDRA \cite{Maplb97}, MSU \cite{Msu}, TAMU \cite{Tamu} , 
EOS \cite{Eos} and GSI \cite{Serfling} etc. 

In addition, the possibility to extract critical exponents and 
critical behavior from finite-size systems has been attempted 
experimentally in \cite{Eos2}.

However, even though the above trials to search for the liquid 
gas phase transition are multi-variant, these signatures are 
still in controversial debating \cite{Camp97,Natowitz,Bauer,Moretto}. 
In this context, it will be helpful and meaningful to present  
novel signature to characterize the nuclear liquid gas phase 
transition to guide the experimental  analysis and 
theoretical predictions. 

In this paper we will introduce, for the first time, Zipf's 
law into the study of nuclear liquid gas phase transition. 
Zipf's law \cite{Crystal} has been known as a statistical 
phenomenon concerning the relation between English words and 
their frequency in literature in the field of linguistics. 
In this Zipf's analysis, one calculates the histogram that 
gives the total number of occurrences of each word in a text. 
If all the words in the text are arranged in rank $n$ order, 
from the most frequent to the least frequent, then such a 
histogram is found to be linear on double logarithmic paper, 
with a slope -$\lambda$, with $\lambda$ $\approx$ 1 for all 
languages. Attempts to understand the origin of the Zipf's 
law are connected to the hierarchical structure of language 
\cite{Mand}. This relation was found not only in linguistics 
but also in other fields of sciences. For examples, the law 
appeared in distributions of populations in cities, 
distributions of areas of lakes, DNA base pair sequences 
\cite{Mant} and cluster-size distribution in percolation 
process \cite{Watanabe}. In this paper, Zipf's law will be 
tested for the charge (or mass)  distribution of nuclear 
clusters and evidenced to be a factor to characterize the 
phase transition. 

Isospin dependent lattice gas model (I-LGM) was used to 
investigate the Zipf's law in nuclear cluster distribution. 
The lattice gas model was developed to describe the liquid-gas 
phase transition for atomic system by Lee and Yang \cite{Yang52} 
first. The same model has already been applied to nuclear 
physics for isospin symmetrical systems in the grancanonical 
ensemble \cite{Biro86} with a sampling of 
the canonical ensemble
 \cite{Camp97,Jpan96,Mull97,Jpan95,Jpan97,Jpan98,Gulm98}, and 
also for isospin asymmetrical nuclear matter in the mean field 
approximation \cite{Sray97}. Here we will make a brief 
description for the I-LGM model.

In the isospin dependent lattice gas  model, $A$ (= $N + Z$) 
nucleons with an occupation number $s$ which is defined $s$ 
= 1 (-1) for a proton (neutron) or $s$ = 0 for a vacancy, are 
placed on the $L$ sites of lattice. Nucleons in the nearest 
neighboring sites interact with an energy 
$\epsilon_{s_i s_j}$. The hamiltonian is written as  
\begin{equation}
E = \sum_{i=1}^{A} \frac{P_i^2}{2m} - 
\sum_{i < j} \epsilon_{s_i s_j}s_i s_j .
\end{equation}
The interaction constant $\epsilon_{s_i s_j}$ is chosen to 
be isospin dependent and be fixed to reproduce the binding 
energy of the nuclei \cite{Jpan98}:
\begin{eqnarray}
 \epsilon_{nn} \ &=&\ \epsilon_{pp} \ = \ 0. MeV \nonumber , \\
 \epsilon_{pn} \ &=&\ - 5.33 MeV,
\end{eqnarray}
 which indicates the repulsion between the nearest neighboring 
like-nucleons and attraction between the nearest neighboring 
unlike-nucleons. This kind of isospin dependent interaction 
incorporates, to a certain extent, Pauli exclusion principle  
and effectively avoids producing unreasonable clusters, such 
as di-proton and di-neutron etc. Three-dimension cubic lattice 
with $L^3$ sites  is used which results in 
$\rho_f$ = $\frac{A}{L^3} \rho_0$ of an assumed freeze-out 
density of disassembling system, in which $\rho_0$ is the 
normal nuclear density. The disassembly of the system is to 
be calculated at $\rho_f$, beyond which nucleons are too far 
apart to interact.  Nucleons are put into lattice by Monte 
Carlo Metropolis sampling. Once the nucleons have been placed 
we also ascribe to each of them a momentum by Monte Carlo 
samplings of Maxwell-Boltzmann distribution. 

Once this is done the I-LGM immediately gives the cluster 
distribution using the rule that two nucleons are part of 
the same cluster if 
\begin{equation}
P_r^2/2\mu - \epsilon_{s_i s_j}s_i s_j < 0.
\end{equation}
 This prescription is evidenced to be similar to the 
Coniglio-Klein's prescription \cite{Coni80} in condensed 
matter physics and shown to be valid in lattice gas type 
calculations \cite{Camp97,Jpan96,Jpan95,Gulm98}.

We first chose the medium size nuclei $^{129}$Xe  to 
analyze the phase transition point behavior in  nuclear 
disassembly 
 and its Zipf's law from the cluster distribution. The 
freeze-out density $\rho_f$ is chosen to be 0.38 $\rho_0$ 
due to the data were best fitted by a $\rho_f$ between 
0.3$\rho_0$ and 0.4$\rho_0$ in the previous LGM 
calculations \cite{Jpan95,Beau96}, which corresponds to  
the cubic lattice  with size $L$ = 7 for $^{129}$Xe. 1000 
events are simulated for each calculation. 

In order to check the phase transition behavior in the I-LGM, 
we will firstly show the results of some physical 
observables, namely the effective power-law parameter, 
$\tau$, the second moment of the cluster distribution, 
$S_2$ \cite{Campi}, and the multiplicity of intermediate 
mass fragments, $N_{imf}$ for the disassembly of $^{129}$Xe 
in figure 1. These observables have been evidenced useful 
in previous works to judge the liquid gas phase transition, 
as shown in Ref. \cite{Jpan98,Maepja,Maprc99}. The valley 
of $\tau$, the peaks of $N_{imf}$ and $S_2$ happens around 
$T$ $\sim$ 5.5 MeV which is the signature of onset of 
phase transition. However, the aim of this paper  is 
searching a novel signature of liquid gas phase transition 
besides the above observables. The above phase transition  
temperature will be only used as a reference of the novel 
signature, as stated below.

Now we  present the results for testing Zipf's law 
in the charge distribution of clusters. The law states 
that the relation between the sizes and their ranks is 
described by $Z_n = c/n$ (n=1, 2, 3, ...), where $c$ is 
a constant and $Z_n$ is the average charge (or mass) of 
rank $n$ in a charge (or mass) list when we arrange the 
clusters in the order of decreasing size. For instance 
the charge  $Z_2$ of the second largest cluster with 
rank $n$ = 2 is one-half of the charge  $Z_1$ of the 
largest cluster, the charge $Z_3$ of the third largest 
cluster with rank $n$ = 3 is one-third of the charge 
$Z_1$ of the largest cluster, and so on. In the 
simulations of this work, we averaged the charges for 
each rank in charge lists of the events: we averaged 
the charges for the largest clusters in each event, 
averaged them for the second largest clusters, averaged 
them for the third largest clusters, and so on. From 
the charges averaged, we examined the relation between 
the charges $Z_n$ and their ranks $n$. Figure 2 shows 
such relations of $Z_n$ and $n$ for Xe in different 
temperature. The histogram is the simulated results 
and the straight lines represent the fit with 
$Z_n \propto  n^{-\lambda}$ in the range of 
1 $\leq$ $n$ $\leq$ 10, where $\lambda$ is the slope 
parameter.  $\lambda$ is 5.77 at $T$ = 3 MeV. Then we 
increased the temperature and examined the same relation  
and obtained $\lambda$ = 3.65 and 1.53 at $T$ = 4 and 
5 MeV, respectively. Up to $T$ = 5.5 MeV, $\lambda$ = 1.00, 
i.e., at this temperature the relation is satisfied to 
the Zipf's law: $Z_n \propto n^{-1} $. When temperature 
continues to increases, $\lambda$ continues to decreases, 
for instance, $\lambda$ = 0.80 at $T$ = 6 MeV and 
$\lambda$ = 0.56 at $T$ = 7.  This temperature having 
the Zipf's law, denoted as $T_A$, is consistent 
with the phase transition temperature obtained in Fig. 1, 
illustrating that the Zipf's law is also a good judgement 
to phase transition. From the statistical point of view, 
the Zipf's law is related to the critical phenomenon 
\cite{Fisher,Stauffer}. Figure 3a summarizes the parameter 
$\lambda$ as a function of temperature. Clearly the 
Zipf's law ($\lambda$ = 1) reveals at phase transition point. 

In order to further illustrate that the Zipf's law 
exists most probably in phase transition point, we directly 
reproduce the histograms with Zipf's law: $Z_n = c/n$. 
In this case, $c$ is sole parameter, but what we are 
interesting in is its truth of the hypothesis of 
Zipf's law: the $\chi^2$ test. Figure 3b demonstrates 
the $\chi^2/ndf$ for the $Z_n$ - $n$ relations at different 
$T$. As expected, there is the minimum $\chi^2/ndf$ around 
the phase transition temperature, which further support that
Zipf's law of the fragment distribution reveals 
when the liquid gas phase transition occurs.

In the above examinations, the system size is finite, 
so the effective phase transition temperature, $T_A$, 
deduced 
from the Zipf's law shall be  only valid for a specific 
finite size nuclei. As well known, the finite size 
scaling is an important factor in studying the phase 
transition of finite matter. In this context, we will 
investigate such an effect for $T_A$ with the help
of the Zipf-type analysis for nuclear clusters and  
predict the value of 
critical temperature, $T_{C}$, at which the Zipf's law 
will appear for the infinite nuclear matter with the 
same $N/Z$ and freeze-out density as $^{129}Xe$ in the 
framework of isospin dependent lattice gas model. 
When we examine the values $T_C$ of thresholds of phase 
transition, we usually extrapolate the effective 
phase transition temperature $T_A$ as 
$L \rightarrow \infty$ \cite{Raynolds}. This is based 
on the extrapolation rule
\begin{equation}
|T_C - T_A| \propto L^{-1/\nu},
\end{equation}
 where $\nu$ is the critical exponent for correlation 
length. For three dimension (3D) percolation class, 
$\nu$ = 0.9 ; for the 3D Ising class or liquid gas class, 
$\nu$ = 0.6 \cite{Muller}. 
Below we will examine if the finite size scaling law  
is valid for the phase transition temperature at which 
the Zipf's law occurs in the nuclear system and tentatively 
extract the critical exponents $\nu$ and $\beta$.

We simulated the nuclear disassembly for systems with 
the same $N/Z$ as $^{129}Xe$ but A = 80, 274, 500, 830 
and 1270 within the cubic sites of 6$^3$, 9$^3$, 11$^3$, 
13$^3$ and $15^3$, respectively, which will result in  
the same freeze-out density as $^{129}Xe$ for comparison. 
The effective phase transition temperatures, $T_A$, at which 
Zipf's law ($\lambda = 1$) appears are extracted  for 
each system.   The figure 4a summarizes such  $T_A$ as 
a function of $L$, from which we can get the two fit 
parameters of  $T_C$ and $\nu$ with the Eq.(4). The 
best fit parameter is shown in the figure, i.e., 
$T_C$ = 5.846 $\pm$ 0.013 MeV and 
$\nu$ = 0.348 $\pm$ 0.023. It tells us that the scaling 
law (Eq.(4)) is rather valid even for the effective 
phase transition temperature deduced from the Zipf's law. 
Viewing from the fit values of $\nu$, it seems to approach  
to  the value of 3D Ising or liquid gas  class than the 3D
percolation class, which  probably 
indicate that the nuclear disassembly belongs to 3D 
liquid gas universality class, 
while the corresponding extrapolated critical temperature 
of $T_C \simeq 5.846$ MeV  for the infinite nuclear matter 
with $N/Z$ = 1.39 and 0.38$\rho_0$ as $A \rightarrow \infty$. 
In addition, the finite size scaling can  be further 
investigated by the source size dependent largest cluster 
at the phase transition point where the Zipf's law appears 
and hence to extract the critical exponent $\beta$ 
by the scaling law:
\begin{equation}
A_{max}/L^3 \propto L^{-\beta/\nu},
\end{equation} 
where $A_{max}$ is the size of the largest cluster at 
the phase transition  point where $\lambda$ = 1  
and $L$ is the lattice size of studied system. For three 
dimension (3D) percolation class, $\beta/\nu$ = 0.44; 
for the Ising class or liquid gas class, 
$\beta/\nu$ = 0.51 \cite{Muller}. 
Figure 4b shows the scaling law of Eq.(5)  for 
A = 80, 129, 274, 500, 830 and 1270. This method 
leads to $\beta/\nu = 1.105 \pm 0.039$ which is close to
the classical value. However, we would like to emphasize that 
one should not be nutty to the value itself of
the exponents because they are very sensitive to the reduced
$T_A$ and $A_{max}$. In addition, the size of lattice  
should be as large as possible if one really wants to extract the 
precise exponents. What we want to say here is that the 
finite size scaling is valid for the Zipf-type analysis 
and it is also possible, in principal, to study the 
critical exponents from there.   

In conclusion,  Zipf-type analysis in the field of 
linguistics is applied to the cluster charge (mass) 
distribution  which has been generated with an isospin 
dependent lattice gas model.  The cluster distributions 
at the phase transition point show exactly inversely to its 
rank, i.e., there exists Zipf's law. This criterion shall 
be useful in searching the nuclear liquid gas phase 
transition experimentally and theoretically in 
combination with the other phase transition observables, 
such as $\tau$, $S_2$ and $N_{imf}$ etc. In addition, 
the finite size effect of the effective phase transition  
temperature at which the Zipf's law appears is also 
addressed and the critical exponents of $\nu$ and 
$\beta$ are tentatively extracted. It seems that the 
nuclear disassembly approaches to the 3D liquid gas 
universality class  in the framework of the isospin 
dependent lattice gas model.  
\\ \\
The author would like to thank Dr. S. Das Gupta and
Dr. J. Pan 
for providing the original LGM code and Dr. B. Tamain
and Dr. O. Lopez for helps. I acknowledge 
 IN2P3-CNRS for financial support of my stay at LPC-Caen 
 during which this work was revised. I also appreciate 
 the members of LPC for the warm hospitality. 
This work was partly  
supported  by the NSFC for Distinguished Young 
Scholar under Grant No. 19725521, the NSFC under 
Grant No. 19705012, the Science and Technology 
Development Fund of Shanghai under Grant 
No. 97QA14038, and the Presidential Foundation of 
Chinese Academy of Sciences.

\figure{Fig.1: The effective power-law parameter, 
$\tau$, the second moment of the cluster distribution, 
$S_2$, and the multiplicity of intermediate mass 
fragments, $N_{imf}$ as a function of temperature 
for the disassembly of $^{129}Xe$. The arrow 
represents the position of phase transition temperature. }

\figure{Fig.2: The average charge $Z_n$ with rank 
$n$ as a function of $n$ for $^{129}Xe$. The 
histograms are the calculation results and the 
straight lines are their fits with 
$Z_n \propto n^{-\lambda}$.}

\figure{Fig.3: The slope parameter $\lambda$ of 
$Z_n$ to $n$ (a) and the $\chi^2$ test for Zipf's 
law (b) as a function of temperature for $^{129}Xe$.  
The arrow represents the position of phase transition
 temperature.}

\figure{Fig.4: The effective phase transition 
 temperature (a) 
and the normalized largest cluster size $A_{max}/L^3$ (b) 
is plotted as a function of  size of lattice. From (a) 
the critical temperature $T_C$ for infinite system and 
the critical exponent $\nu$ can be extracted; From (b) 
the critical exponent $\beta/\nu$ can be  obtained.}

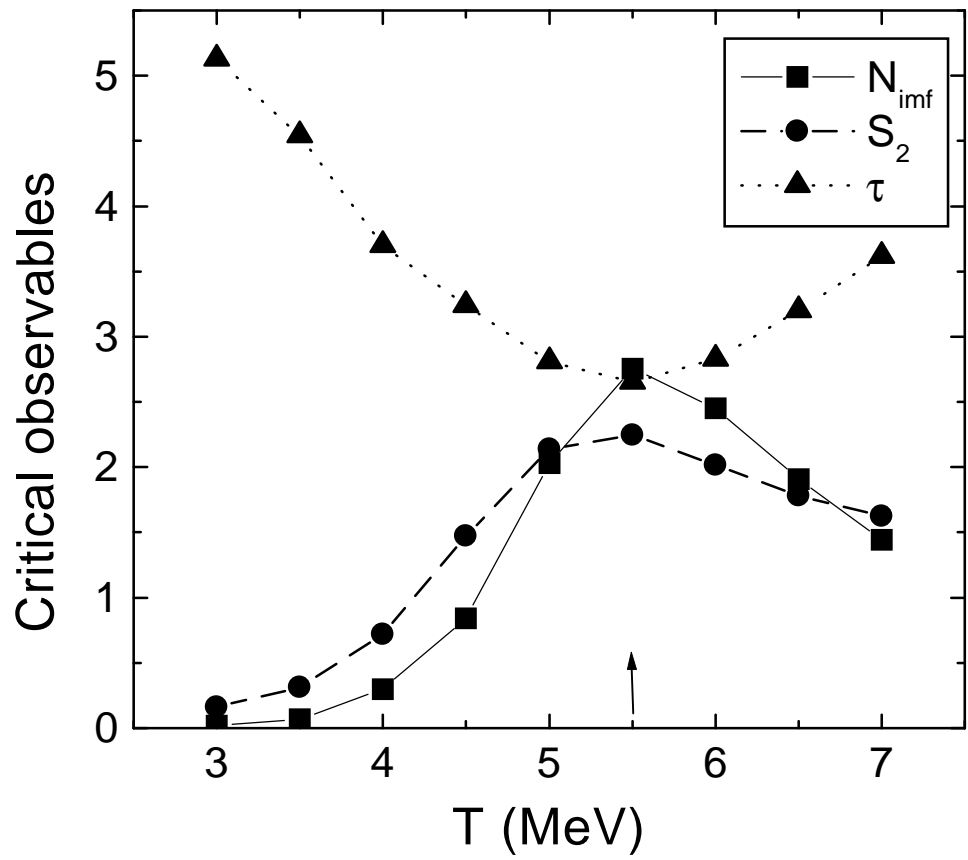
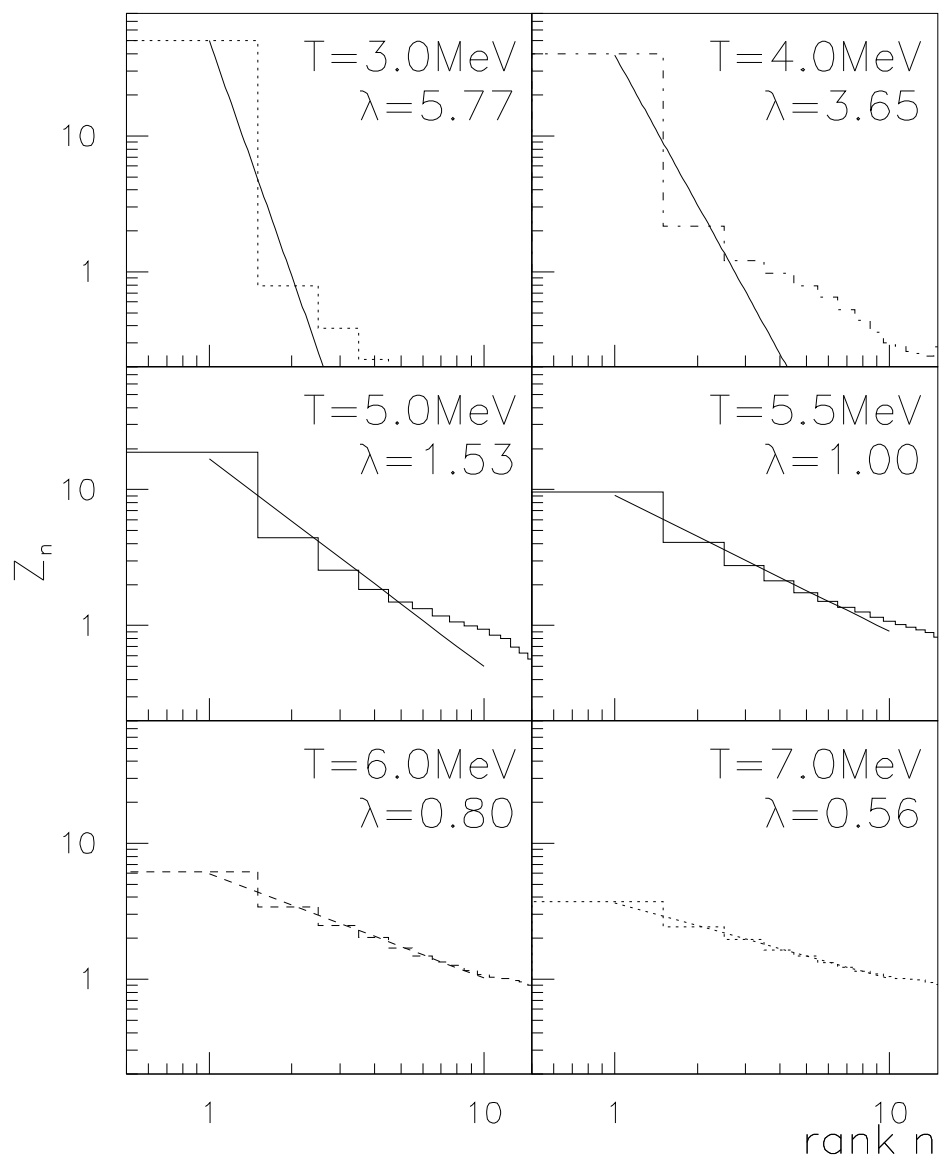
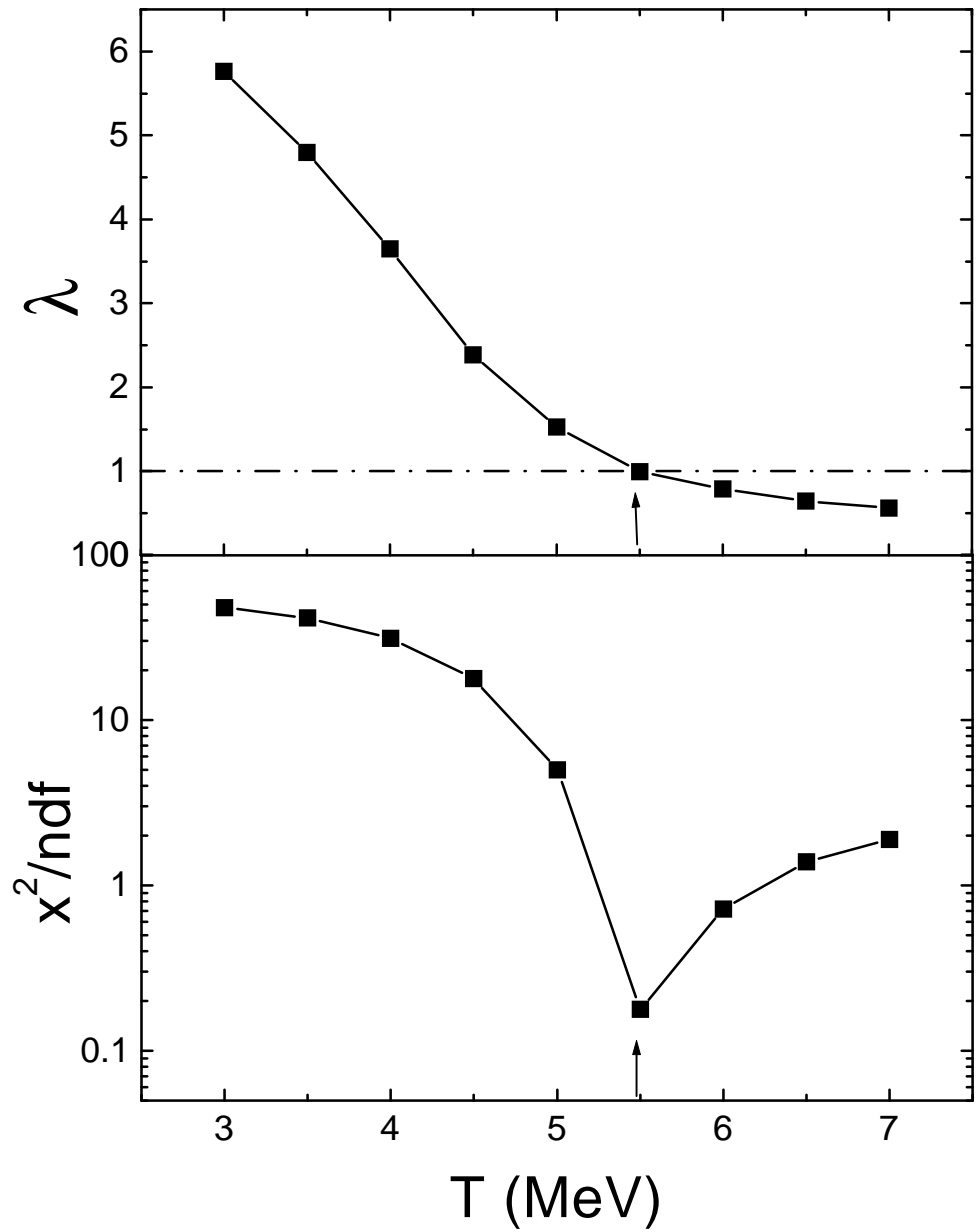
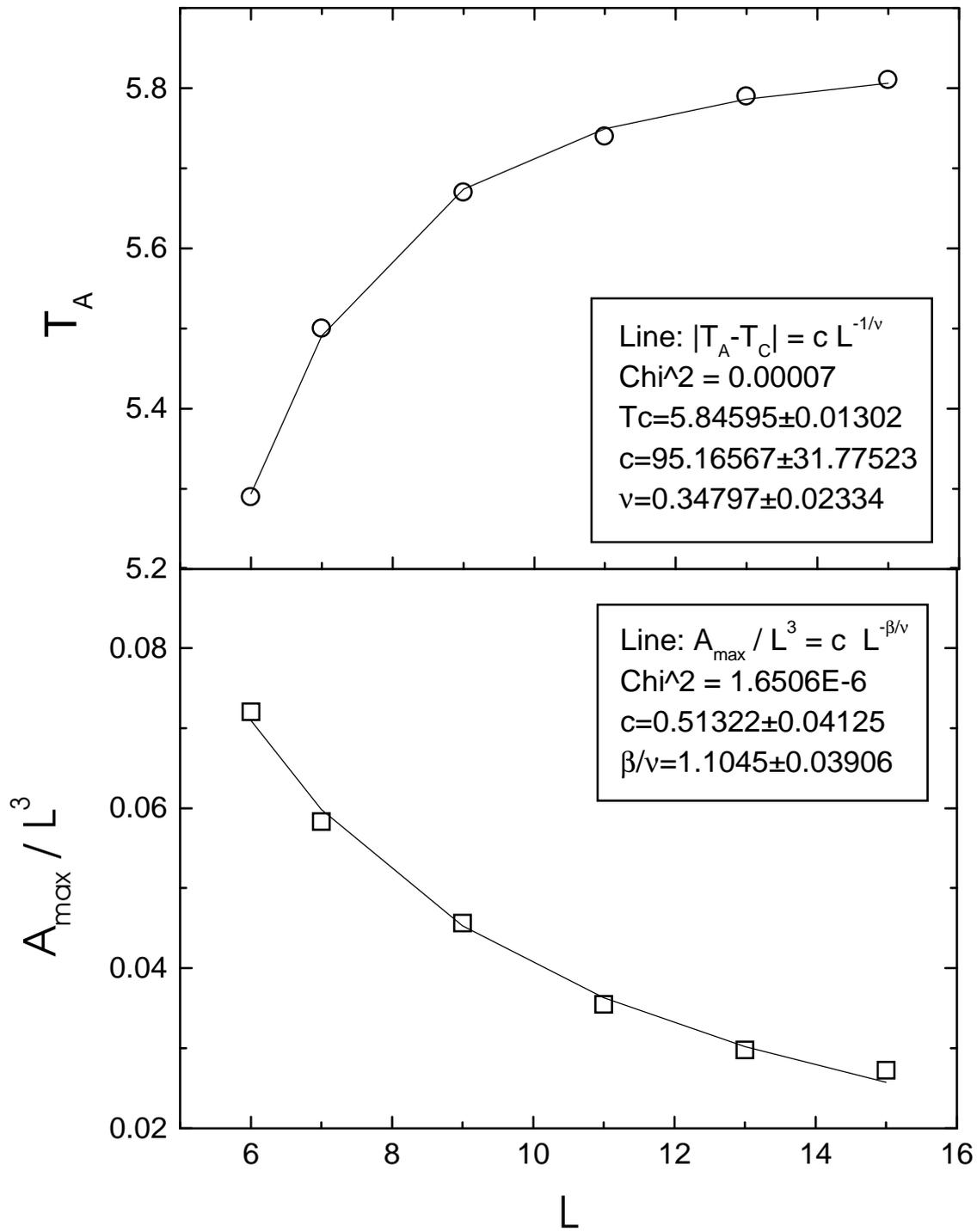
\end{document}